\documentclass[review]{elsarticle}
\pdfoutput=1
\usepackage{lineno,hyperref}
\modulolinenumbers[5]

\journal{Journal of Control Engineering Practice}

\usepackage{AHGhasemi}
\usepackage{mathrsfs}
\usepackage{amsmath}
\usepackage{amsfonts}
\usepackage{amssymb,latexsym}
\usepackage{epsfig}
\usepackage{psfrag}
\usepackage{indentfirst}
\usepackage{appendix}
\usepackage{epstopdf}
\usepackage{enumerate}
\usepackage{tikz}
\usepackage{caption} 
\usepackage{picture}
\usepackage[ruled,vlined]{algorithm2e}
\usepackage{multirow}
\usepackage{natbib}

\usepackage{color, colortbl}
\definecolor{Gray}{gray}{0.9}
\definecolor{black}{rgb}{0.88,1,1}

\usepackage{enumitem}                                
\usepackage{todonotes}

\usepackage{multicol}

\begin{document}

\begin{frontmatter}

\title{Modulation of Control Authority  in  Adaptive Haptic Shared Control Paradigms}

\author[mymainaddress]{Vahid Izadi}

\author[mymainaddress]{Amir H. Ghasemi\corref{mycorrespondingauthor}}
\cortext[mycorrespondingauthor]{Corresponding author}
\ead{ah.ghasemi@uncc.edu}

\address[mymainaddress]{Department of Mechanical Engineering and Engineering Science,  University of North Carolina Charlotte, Charlotte, NC, 28223}

\begin{abstract}
This paper presents an adaptive haptic shared control framework wherein a driver and an automation system are physically connected through a motorized steering wheel. The automation system is modeled as an intelligent agent that is not only capable of making decisions but also monitoring the human's behavior and adjusting its behavior accordingly. 
To enable the automation system to smoothly exchange the control authority with the human partner, this paper introduces a novel self-regulating impedance controller for the automation system. To determine an optimal modulation policy, a cost function is defined. The terms of the cost function are assigned to minimize the performance error and reduce the disagreement between the human and automation system. To solve the optimal control problem, we employed a nonlinear model predictive approach and used the continuation generalized minimum residual method to solve the nonlinear cost function. To demonstrate the effectiveness of the proposed approach, simulation studies consider a scenario where the human and the automation system both detect an obstacle and negotiate on controlling the steering wheel so that the obstacle can be avoided safely. The simulations involve four interaction modes addressing the cooperation status (cooperative and uncooperative) and the desired direction of the control transfer (active safety and autopilot). The results of the numerical studies show that when the automation system acts as autopilot, using the proposed modulation method, the automation adopts smaller impedance controller gains, which results in a smaller disagreement between the human and automation systems. On the other hand, when the human's control command is insufficient,  by modulating and adopting larger values for the impedance controller parameters, the automation system gains the control authority and ensures the safety of the obstacle avoidance task.

\end{abstract}

\end{frontmatter}


\section{Introduction}

Haptic shared control paradigms have a wide range of applications from transformative technologies in which a fully autonomous system is not yet accessible/feasible (e.g., service robots, semi-autonomous vehicles, smart manufacturing) to applications where human-robot interactions are inevitable, or even desirable (e.g., rehabilitative devices, care robots, and educational robots) \citep{agah2000human, albu2005physical, beyl2011safe, boehm2016architectures, ghasemi2016role,ghasemi2018adaptive, ghasemi2018game,ghasemi2019shared, vitiello2013neuroexos, bhardwaj2020s, izadi2019determination}.    In a haptic shared control paradigm, both humans and co-robots can simultaneously exert their control inputs, and by virtue of haptic feedback {continuously,} monitors each other's actions.  Traditionally, haptic shared control paradigms consist of interactive robots that were designed to act mainly as reactive followers where the robot (with some level of autonomy) followed the human's commands \cite{ gillespie1998virtual, haanpaa1997advanced, park2001virtual}. However, this type of master-servant arrangement does not capture the sense of partnership \cite{ ikeura2002optimal, ikeura1997variable, arai2000human} that we mean when we speak of two humans cooperatively moving a piece of furniture.  Nowadays, with recent advancements in robotics and artificial intelligence, a co-robot as a pro-active partner can be designed to monitor human actions, communicate its behavior, and even communicate and exchange roles with a human partner\cite{reed2008physical, groten2009experimental}.


To enable co-robots to comprehend how human behaviors change during the interaction, a wide range of research has been devoted to human-human interaction to learn from the behavioral mechanisms that humans utilize. For instance, Reed and Peshkin studied human-human interaction and recognize different humans select different patterns for interaction \cite{reed2008physical}.  Stefanov et al. defined two roles of conductor and executor in the execution of a haptic task \cite{stefanov2010online}.   Orguz et al. proposed a haptic negotiation framework to address the role exchange in a dynamic task \cite{oguz2010haptic, oguz2012supporting}. In these preceding studies, it was shown that the role arbitration could dynamically change between the human and co-robot during task execution. Specifically, the role exchanges aim to either increase the robot’s autonomy level at the expense of the human’s authority; we call this interaction mode the active safety mode; or, conversely, increase the human’s control over the shared activity at the expense of the robot’s autonomy;  we call this interaction mode the autopilot mode. 
However, it is vital for an intuitive role arbitration that the co-robot smoothly transfer the control authority \cite{sheridan2011adaptive, inagaki2003adaptive}.
To support smooth transfers of authority and harness the complementary features of human and automatic control, many of the efforts in this area has been devoted to the development of standard impedance control \cite{lee2014upper}, force control \cite{righetti2014autonomous},
or hybrid interaction controllers \cite{adhikary2017hybrid,anderson1988hybrid} for the co-robot. The quasi-static performance of these controllers is highly dependent on the choice of parameters, i.e., stiffness and damping \cite{balatti2018self}. In most cases, such parameters are preset,  which limits the adaptation capability of co-robots to varying interaction modes. To improve the customizability feature of co-robots and enable them to naturally interact with different humans through a haptic interface and personalize their interactions based on recognizing the human's preference in real-time, this paper introduces a novel self-regulating impedance controller.

To this end, we present the co-robot with a similar structure to the human partner. In particular, we consider the co-robot as a two-level hierarchical control structure. While the higher-level controller generates the co-robot desired reference (intent), the lower level is an impedance controller which its output is the co-robot’s control force/torque. 
We define a cost function to determine how co-robot impedance controller gains should be dynamically modulated so that a smooth transition of control authority can occur. The cost function maximizes task performance while minimizing the disagreement between the human and co-robot within different interaction modes. To solve the optimal control problem,  we utilize a nonlinear model-predictive control approach.  Specifically, we employed the continuation generalized minimum residual (C/GMRES) solver that provides an iterative algorithm to solve the nonlinear model predictive controller \cite{mori2007continuation,soneda2005nonlinear,ohtsuka1994stabilized,ohtsuka2000continuation}.  In this method, first, the optimal control problem is discretized over the horizon. A differential equation is then obtained through the use of the continuation method to update the sequence of control inputs \cite{ohtsuka2000continuation}. Since the differential equation involves a large linear equation, the GMRES  method \cite{kelley1995iterative} is employed to solve the linear equation. It is shown that the C/GMRES requires much less computational expenses than other iterative methods such as Newton’s method. Moreover, C/GMRES involves no line search, which is also a significant difference from standard optimization methods \cite{ohtsuka2004continuation}. 

While the fundamental approaches and models proposed in this research can be applied to a wide range of physical human robot systems, we select steering control of semi-automated vehicles as a setup for exploring the proposed study. We consider a scenario where the human and the automation system detect an obstacle and negotiate on controlling the steering wheel so that the obstacle can be avoided safely.  To this end, the simulations involve four interaction modes addressing the cooperation status (cooperative and uncooperative) and the desired direction of the control transfer (active safety and autopilot).  

The outline of this paper is as follows. In section \ref{modelingsec}, we model the adaptive haptic shared control paradigm, and the equations of motions are derived. In section \ref{controlsec}, the problem of modulation of control authority is presented as an optimal control problem. Specifically, we describe how the C/GMRES method is used to determine an optimal modulation policy that allows a smooth transition of control authority. To demonstrate the effectiveness of the proposed approach, a set of numerical simulations are illustrated in Section 4. Specifically, based on the cooperation status (cooperative and uncooperative) and the desired direction of the control transfer (to automation or human), four interaction scenarios are defined. The numerical results compare the performance of the proposed adaptive haptic shared control paradigm with a non-adaptive haptic shared control paradigm in the steering control task. Section 5 consists of the conclusions and the future directions for this research.

\section{Adaptive Haptic Shared Control Framework \label{modelingsec}} 
\begin{figure}[htbp]
    \centering
    \includegraphics[width=\textwidth]{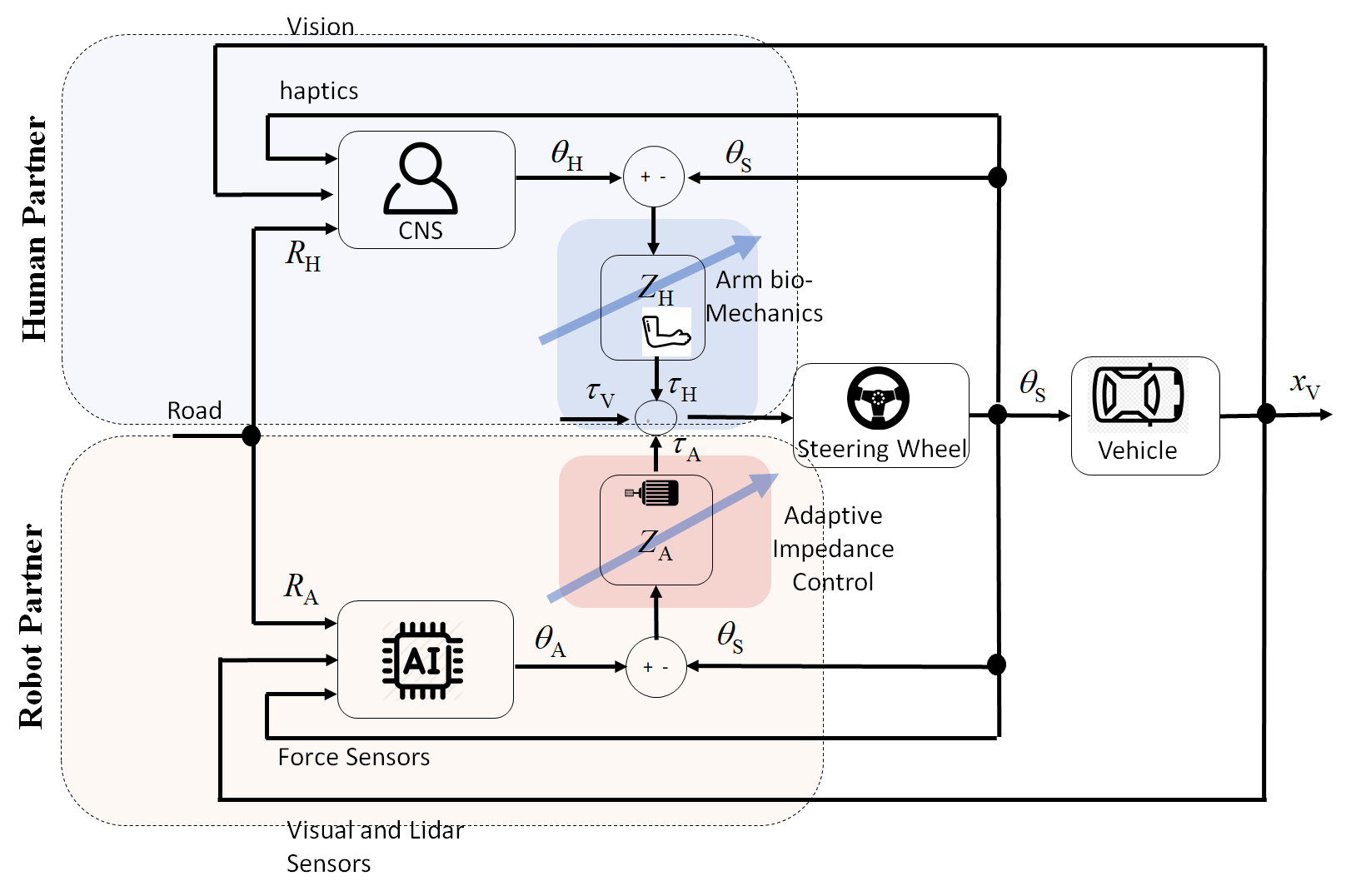}
    \caption{ A general model of control sharing between driver and automation. 
    }
    \label{HRI}
\end{figure}

Figure \ref{HRI} shows a  schematic of an adaptive haptic shared control paradigm.  Three entities each impose a torque on the steering wheel: a driver through his hands, an automation system through a motor, and the road through the steering linkage.  

In this paper, we model the human and automation system with a similar structure. In particular, we model the driver as a hierarchical two-level controller. The upper-level control represents the cognitive controller, and its output,  $\theta_{\rm H}$, represents the driver’s intent. 
The lower-level represent the human's biomechanics, $Z_{\rm H}$, and is considered back-drivable \cite{boehm2016architectures}. To indicate that driver's biomechanic parameters vary with changes in grip on the steering wheel, use of one hand or two, muscle co-contraction, or posture changes, we have drawn an arrow through human $Z_{\rm H}$.

Similarly, the automation system is modeled as a higher-level controller (AI) coupled with a lower-level impedance controller. The automation system is also considered to be back-drivable, and the gains of the impedance controller, $Z_{\rm A}$, are designed to be modest rather than infinite. In other words, the automation is not intended to behave as an ideal torque source; instead, the automation imposes its command torque $\tau_{\rm A}$ through an impedance $Z_{\rm A}$ that is approximately matched to the human impedance $Z_{\rm H}$.

 Furthermore, the reference signals $R_{\rm H}$ and $R_{\rm A}$ represent the goals of the driver and the automation system, respectively. It should be noted that these goals may not necessarily be the same, which is when the negotiation of control authority becomes essential.  To generate algorithms that support the negotiation and dynamic transfer of the control authority between the human and co-robot, the robot can adjust its behavior at a higher level (changing intent) as well as in the lower level (changing $Z_{\rm A}$). Specifically,  from the model presented in Figure \ref{HRI}, it follows that the steering angle  $\theta_{\rm S}$ is not only a function of the human’s intent $\theta_{\rm H}$, automation’s intent $\theta_{\rm A}$, and the road feedback torque $\tau_{\rm V}$, but also its a function of human arm's biomechanics  $Z_{\rm H}$ as well as the gains of the impedance controller $Z_{\rm A}$ \cite{bhardwaj2019estimating}.
 The crux of this paper lies in the design of a back-drivable impedance $Z_{\rm A}$ such that it enhances the negotiation and transfer of control authority between the human and automation system. To this end, we present the equations of motion of the lower-level of the adaptive haptic shared control framework shown in Figure \ref{EoM}.

\subsection{Equations of Motion}

\begin{figure}[htbp]
    \centering
    \includegraphics[width=\textwidth]{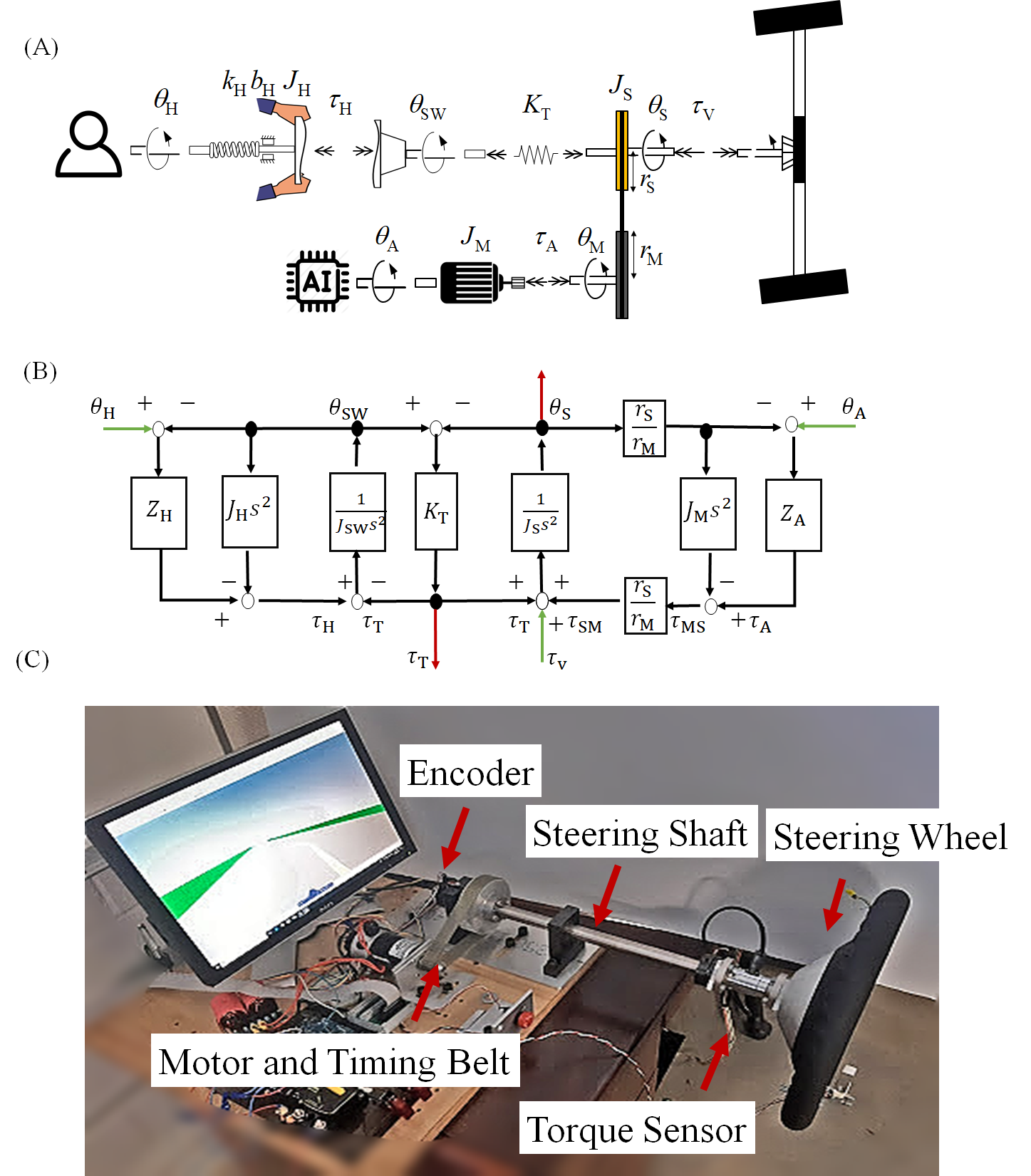}
    \caption{(A) Torque/intent block diagram for the general model of the shared driving control, (B) A block diagram is laid out to highlight the interaction ports between subsystems, (C) Experimental Platform
    }
    \label{EoM}
\end{figure}

Figure \ref{EoM}-A shows a free body diagram of an adaptive haptic shared control paradigm consisting of a driver, a steering wheel, a steering shaft, and an automation system. Figure \ref{EoM}-A demonstrates a simplified model of a driver arms' biomechanics in the form of a mass-spring-damper system connected to a mass-less cart representing the driver's intent, $\theta_{\rm H}$. The steering wheel is modeled as a disk with a rotational inertia of $J_{\rm SW}$.  A differential torque sensor is modeled as a rotational spring with stiffness $K_{\rm T}$ and connected to the steering wheel and steering shaft. The steering shaft is also considered as a rotational bar with the inertia of $J_{\rm S}$ that is connected to the steering wheel on the left side,  to the rack and pinion on the right side and the automation system through a timing belt with a mechanical advantage of $r_{\rm S}/r_{\rm M}$. The block diagram of the lower-level of the adaptive haptic shared control is also shown in Figure \ref{EoM}-B. In this block diagram, three signals of $\theta_{\rm H}, \theta_{\rm A}$, and $\tau_{\rm V}$ are considered as exogenous signals; and two signals of differential torque $\tau_{\rm T}$ and the steering shaft angle $\theta_{\rm S}$ are measured. Figure \ref{EoM}-C shows an apparatus of a haptic steering wheel that can be used for implementing the proposed shared control scheme. To ensure the back-drivability of the automation system,  a  motor with a low-inertia and a pulley with a low mechanical advantage should be selected.  The steering wheel is equipped with a torque transducer to measure the differential torque between the human and robot, and an encoder for measuring the steering column position and velocity.


It follows from Figure \ref{EoM} that the equations of motion for the steering wheel, steering column and the motor can be expressed as 
\begin{align}
    &J_{\rm SW} \ddot \theta_{\rm SW}=\tau_{\rm H}-\tau_{\rm T} \label{TethaSW} \\
    &J_{\rm S} \ddot \theta_{\rm S}=\tau_{\rm T}+\tau_{\rm V}+\tau_{\rm SM}\label{TethaS} \\
    &J_{\rm M}\ddot \theta_{\rm M}=\tau_{\rm A}-\tau_{\rm MS}\label{TethaM}
\end{align}
where $\tau_{\rm SM}$ and $\tau_{\rm MS}$ represent the internal torque imposed by the timing belt. It should be noted that the kinematic and kinetic constraints imposed by the timing belt are $r_{\rm M} \theta_{\rm M}=r_{\rm S} \theta_{\rm S}$ and $r_{\rm M} \tau_{\rm SM}=r_{\rm S} \tau_{\rm MS}$.

Modeling the driver as a spring-mass-damper with a proximal motion source $\theta_{\rm H}(t)$, the torque applied by the human is \cite{ghasemi2018adaptive}
\begin{align}
    \tau_{\rm H}=-J_{\rm H}\ddot \theta_{\rm SW}+B_{\rm H}(\dot \theta_{\rm H}-\dot \theta_{\rm SW})+K_{\rm H}( \theta_{\rm H}- \theta_{\rm SW}) \label{tauH}
\end{align}
where $J_{\rm H}$,  $ B_{\rm H}$, and $ K_{\rm H}$ are  the inertia, damping and stiffness of the driver's arm. Similarly, considering an impedance controller  in the lower-level of the automation system, the toque generated by the motor can be presented as 
\begin{align}
    \tau_{\rm A}=&B_{\rm A}(\dot \theta_{\rm A}-\dot \theta_{\rm M})+K_{\rm A}( \theta_{\rm A}- \theta_{\rm M})
    \nn \\=&B_{\rm A}(\dot \theta_{\rm A}-\frac{r_{\rm S}}{r_{\rm M}}\dot \theta_{\rm s})+K_{\rm A}( \theta_{\rm A}- \frac{r_{\rm S}}{r_{\rm M}}\theta_{\rm S}) \label{tauA}
\end{align}
where $K_{\rm A}$, $B_{\rm A}$ represent the gains of the impedance controller. Furthermore, it follows from Figure \ref{EoM} that the torque measured by the torque sensor can be expressed as 
\begin{align}
    \tau_{\rm T}=K_{\rm T} (\theta_{\rm SW}-\theta_{\rm S}) \label{tauT}
\end{align}

It follows from Eq. (\ref{tauH}) that human's torque is not only a function of human's intent $\theta_{\rm H}$ but also the biomechanic parameters $Z_{\rm H}$. By modulating these parameters, the human can either yield or retain the control authority. Similarly, algorithms can be developed to give the automation the ability to either yield authority or retain authority as a function of driver behavior and sensed threats to safety.  To present how human's bio-mechanics and the automation's impedance controller parameters may evolve in time, we introduce  the following simple but generic dynamic models as
\begin{align}
    \dot Z_{\rm H}(t)&=\alpha_{\rm H} Z_{\rm H}(t)+\beta_{\rm H}\Gamma_{\rm H}(t) \label{ZH}\\
    \dot Z_{\rm A}(t)& =\alpha_{\rm A} Z_{\rm A}(t)+\beta_{\rm A}\Gamma_{\rm A}(t) \label{ZA}
\end{align}
where $Z_{\rm H}=[B_{\rm H} \ K_{\rm H} ]^{\rm T}$, $Z_{\rm A}=[B_{\rm A} \ K_{\rm A} ]^{\rm T}$, and $\Gamma_{\rm H}=[ \Gamma_{\rm bH}(t)\ \Gamma_{\rm kH}(t)]^{\rm T}$ is the human’s control action for modulating his impedance $Z_{\rm H}$ and $\Gamma_{\rm A}=[ \Gamma_{\rm bA}(t)\ \Gamma_{\rm kA}(t)]^{\rm T}$ is the automation’s control input for modulating its impedance $Z_{\rm A}$. Additionally, 
\begin{align}\label{al_bet}
\alpha_{\rm H}=&\left[ \begin{matrix}
            \alpha_{\rm bH} & 0 \\ 0 & \alpha_{\rm kH}\end{matrix}\right], \quad 
            \beta_{\rm H}= \left[ \begin{matrix}
            \beta_{\rm bH} & 0 \\ 0 & \beta_{\rm kH}\end{matrix}\right] \nn 
            \\
          \alpha_{\rm A}=&  \left[ \begin{matrix}
            \alpha_{\rm bA} & 0 \\ 0 & \alpha_{\rm kA}\end{matrix}\right],
   \quad \beta_{\rm A}= \left[ \begin{matrix}
            \beta_{\rm bA} & 0 \\ 0 & \beta_{\rm kA}\end{matrix}\right] \nn
\end{align}
where $\{\alpha_{\rm bH}, \alpha_{\rm kH}, \alpha_{\rm bA}, \alpha_{\rm kA}, \beta_{\rm bH}, \beta_{\rm kH}, \beta_{\rm bA}, \beta_{\rm kA}\}$ are constant parameters.

Ideally, to determine an optimal behavior for the automation system, optimization should be performed over all control signals of the automation system, including (i.e., $\theta_{\rm A}, \Gamma_{\rm A}$). However, the focus of this paper is to determine $\Gamma_{\rm A}$  as a means for allocating the level of authority between the driver and the automation system.

By combining  Eqs. (\ref{TethaSW}-\ref{ZA}), the  dynamics  interaction between  human and automation system in the lower-level of the adaptive haptic shared control framework  can be expressed as 
\begin{align}
    \dot x(t)&=f(x(t),w(t))+Bu(t), \\ y(t)&=h(x(t)) 
      \label{Nonlinear_State-space}
\end{align}
where $x=[\theta_{\rm SW}\ \dot \theta_{\rm SW}\ \theta_{\rm S}\ \dot \theta_{\rm S}\ B_{\rm H}\ K_{\rm H}\ B_{\rm A}\ K_{\rm A}]^{\rm T}$, are the state of the system; $u=[  \Gamma_{\rm bA}(t)\ \Gamma_{\rm kA}(t)]^{\rm T}$ are the control commands, and  $w=[\Gamma_{\rm bH}(t) \ \Gamma_{\rm kH}(t) \ \theta_{\rm H} \ \theta_{\rm A} \  \tau_{\rm V}]^{\rm T}$ are the exogenous signals, $y=[\theta_{\rm S}\ \tau_{\rm T}\ K_{\rm H}\ B_{\rm H}\ K_{\rm A}\ B_{\rm A} ]^{\rm T}$  are measured variables, and 
\begin{align}
    f\left(x,w\right)=\left[\begin{matrix}{\dot{\theta}}_{\rm SW}\\\frac{B_{\rm H}\left({\dot{\theta}}_{\rm H}-{\dot{\theta}}_{\rm SW}\right)+K_{\rm H}\left(\theta_{\rm H}-\theta_{\rm SW}\right)-K_{\rm T}\left(\theta_{\rm SW}-\theta_{\rm S}\right)}{J_{\rm SW}+J_{\rm H}}\\{\dot{\theta}}_{\rm S}\\\frac{\left(\frac{r_{\rm S}}{r_{\rm M}}B_{\rm A}\left({\dot{\theta}}_{\rm A}-\frac{r_{\rm S}}{r_{\rm M}}{\dot{\theta}}_{\rm S}\right)+\frac{r_{\rm S}}{r_{\rm M}}K_{\rm A}\left(\theta_{\rm A}-\frac{r_{\rm S}}{r_{\rm M}}\theta_{\rm S}\right)+K_{\rm T}\left(\theta_{\rm SW}-\theta_{\rm S}\right)+\tau_{\rm v}\right)}{J_{\rm S}+\left(\frac{r_{\rm S}}{r_{\rm M}}\right)^2J_{\rm M}}\\{\alpha_{\rm bH}B}_{\rm H}+\beta_{\rm bH}\Gamma_{\rm bH}\\\alpha_{\rm kH}K_{\rm H}+\beta_{\rm kH}\Gamma_{\rm kH}\\\alpha_{\rm bA}B_{\rm A}\\\alpha_{\rm kA}K_{\rm A}\\\end{matrix}\right],
    B=\left[\begin{matrix} 0 & 0\\ 0&0 \\0&0 \\0&0 \\0 & 0\\0 & 0\\\beta_{\rm bA} & 0 \\0  &\beta_{\rm kA}
    \end{matrix}\right]
\end{align}

\section{Impedance Modulation Controller Design \label{controlsec}}

In this section, we present a predictive controller that is used to modulate the automation's impedance controller parameters to enhance the assistive behavior of the automation system. For the steering control problem, we  define a nonlinear cost function $J(t)$ in the form of 
\begin{align}
    \min_{\Gamma_{\rm A}}  J(t) = \int_{t}^{t+t_h}\{ \|\theta_{\rm H}(t) - \theta_{\rm S}(t)   \|_{w_1}+\|\theta_{\rm A}(t) - \theta_{\rm S}(t)   \|_{w_2} +\|\tau_{\rm T}(t)\|_{w_3}\}\label{Cost_func_org}
\end{align}
where $t_h$ is the defined horizon for the model predictive controller, $w_1$, $w_2$ and $w_3$ are weights  matrices. The first term of the cost function aims to minimize the error between the human's intent and the steering angle. Similarly, the second term of the cost function is defined to minimize the tracking error between the automation's desired angle (automation's intent) and the steering angle. Since the human's and automation's intent may not necessarily be the same, which is when the negotiation of control authority becomes important, the third term of the cost function is defined to minimize the disagreement between a driver and the automation system. 

We define two sets of constraints for the nonlinear cost function $J$ to ensure the non-negative values for the gains of the impedance controller. In particular,  
\begin{align}
    C_1(t) : \{ s_1^2 - {B}_{\rm A}(t) = 0, {B}_{\rm A}(t) & \geqslant{0} \}
    \label{Constraints_ineq_1}\\
    C_2(t) : \{ s_2^2 - {K}_{\rm A}(t) = 0, {K}_{\rm A}(t) & \geqslant{0} \}
    \label{Constraints_ineq_2}
\end{align}
where $s_1$ and $s_2$ are slack variables. By using the non-negative slack variables in Eqs. (\ref{Constraints_ineq_1}) and (\ref{Constraints_ineq_2}), the inequality constraints will be transformed to the equality constraints \cite{tapia1979role}.

To solve the nonlinear cost function described in Eq. \ref{Cost_func_org}, we discretize the equation of the dynamics system using the forward Euler method.
Specifically, 
\begin{align}
    x^{(k+1)} =  x^{(k)} + T_{\rm s} f \left ( x^{(k)}, w^{(k)} \right )+T_{\rm s} B u^{(k)}
    \label{Discretized_State-space}
\end{align}
where $T_{\rm s}$ is the size of the time-step, $k$ is the number of time-step (considered as the current time-step), $x^{(k)}$, $w^{(k)}$ and $u^{(k)}$ are equal to $x{(t = T_{\rm s} k)}$, $w(t = T_{\rm s} k)$ and $u(t = T_{\rm s} k)$,  respectively. It should be noted that that higher order discretizations can be employed  at the expense of the computational complexity. Furthermore, the cost function $J$ and the constraints $C_1$ and $C_2$ can be discretized as 
\begin{align}
     &\min_{\Gamma_{\rm A}}  J^{(k)} = \sum_{j=1}^{N_{\rm p}}T_{\rm s}\{\|\theta_{\rm H}^{(k+j)} - \theta_{\rm S}^{(k+j)}   \|_{w_1}+\|\theta_{\rm A}^{(k+j)} - \theta_{\rm S}^{(k+j)}   \|_{w_2}+\|\tau_{\rm T}^{(k+j)}\|_{w_3}\} \nonumber\\ 
    & \rm{subjected} \quad  \rm{to}: \quad  \left \{\begin{matrix}
x^{(k+1)} =  x^{(k)} + T_{\rm s} f \left ( x^{(k)}, w^{(k)}   \right )+T_{\rm s} B u^{(k)}\\ \\
C_1^{(k)}: \{ s_1^2 - {B}_{\rm A}^{(k)} = 0, \quad {B}_{\rm A}^{(k)}  \geqslant{0} \}\\ \\
C_2^{(k)}: \{ s_2^2 - {K}_{\rm A}^{(k)} = 0, \quad{K}_{\rm A}^{(k)}  \geqslant{0} \}
\end{matrix} 
    \right.
    \label{Disc_Problem}
\end{align}
Next, Let $H$ denote the Hamiltonian defined by

\begin{align}
    H\left(x^{(k)}, w^{(k)},u^{(k)}, \lambda^{(k)}, \mu^{(k)} \right) &= T_{\rm s}\left( \|\theta_{\rm H}^{(k)} - \theta_{\rm S}^{(k)}   \|_{w_1}+\|\theta_{\rm A}^{(k)} - \theta_{\rm S}^{(k)}   \|_{w_2}+\|\tau_{\rm T}^{(k)}\|_{w_3}\right)\nonumber \\
    &+\lambda^{(k)} \left (x^{(k)}-x^{(k+1)}+T_{\rm s} f \left ( x^{(k)}, w^{(k)} \right )+T_{\rm s} B u^{(k)}\right )\nonumber \\
    &+\mu ^{(k)}\left ( {\left[C_1^{(k)}, C_2^{(k)}\right]}^{\rm T}\right)
    \label{Lagrangian}
\end{align}
where
\begin{align}
&\lambda^{(k)} = \left [ \begin{matrix}
\lambda_{\theta_{\rm{S}}}^{(k)} \quad  \lambda_{\dot{\theta}_{\rm{S}}}^{(k)} \quad 
\lambda_{\theta_{\rm{SW}}}^{(k)} \quad \lambda_{\dot{\theta}_{\rm{SW}}}^{(k)} \quad 
\lambda_{\Gamma_{\rm bH}}^{(k)} \quad
\lambda_{\Gamma_{\rm kH}}^{(k)}\quad
\lambda_{\Gamma_{\rm bA}}^{(k)} \quad
\lambda_{\Gamma_{\rm kA}}^{(k)} \end{matrix} \right ]\\
&\mu^{(k)} = \left [ \begin{matrix}
\mu^{(k)}_{C_{1}}&& \mu^{(k)}_{C_{2}}
\end{matrix} \right ]
    \label{Lagrangian_var}
\end{align}
where $\lambda$ and $\mu$ are costate vector and Lagrange multiplier vector respectively. 
Next, we  construct the discrete Lagrangian function  as
\begin{align}
\mathcal{L} (X,U) &= \sum_{j=1}^{N_{\rm p}}H\left(x^{(k+j)}, w^{(k+j)},u^{(k+j)}, \lambda^{(k+j)}, \mu^{(k+j)} \right)
\end{align}
where vectors $X \in \mathbb R^{11\times N_{\rm p}}$ and $U \in \mathbb R^{12\times N_{\rm p}}$ are
\begin{align}
&X = [x^{(k)}, w^{(k)}, x^{(k+1)}, w^{(k+1)}, ..., x^{(k+N_{\rm p})}, w^{(k+N_{\rm p})}]^{\rm T}\nonumber\\ 
&U = [u^{(k)}, \mu^{(k)}, \lambda^{(k)},\cdots, u^{(k+N_{\rm c})}, \mu^{(k+N_{\rm c})}, \lambda^{(k+N_{\rm c})}, \cdots, u^{(k+N_{\rm c})}, \mu^{(k+N_{\rm p})}, \lambda^{(k+N_{\rm p})}]^{\rm T} \nonumber
\end{align}
where $N_{\rm c}$ is the control horizon. Note that for $N_{\rm c}\le j\le N_{\rm p}$, $u^{k+j}=u^{k+N_{\rm c}}$. We also define a projection matrix $P_0$ as 
\begin{align}
    P_0=\left[\begin{matrix} 
    1 & 0 & 0& \cdots & 0 & 0 & 0\\
    0 & 1 & 0& \cdots & 0 & 0 & 0
    \end{matrix}\right]_{2 \times 12N_{\rm p}}.
\end{align}

It should be mentioned that by employing forward recursion the state variables $x^{(k+j)}$, $j = 1, \cdots, N_{\rm p}$ can be defined using the system dynamics. Furthermore,  the co-states $\lambda^{(k+j)}$ and Lagrange multipliers $\mu^{(k+j)}$ can be determined employing back recursion from the final condition to the present time-step ($j = N_{\rm p}, N_{\rm p}-1, \cdots, 1$).

By applying the Karush–Kuhn–Tucker (KKT) first order necessary condition, the solution of nonlinear equations $\frac{\partial H}{\partial \lambda} = 0$, $\frac{\partial H}{\partial u} = 0$ and $\frac{\partial H}{\partial \mu} = 0$ construct the candidate optimal points. Therefore, to determine the optimal parameters of the impedance controller, the first order KKT vector $F\left(X, U, t\right)$ for $N_{\rm p}$ horizon can be defined as 
\begin{align}
F\left(X, U, t\right)= 
\left [ \begin{matrix} 
\frac{\partial H^{\rm T}\left(x^{(k)}, w^{(k)}, u^{(k)}, \lambda^{(k)}, \mu^{(k)}\right)}{\partial u} \\ \\
\frac{\partial H^{\rm T}\left(x^{(k)}, w^{(k)}, u^{(k)}, \lambda^{(k)}, \mu^{(k)}\right)}{\partial \lambda}\\ \\
\frac{\partial H^{\rm T}\left(x^{(k)},  w^{(k)}, u^{(k)}, \lambda^{(k)}, \mu^{(k)}\right)}{\partial \mu}\\ 
\vdots\\
\frac{\partial H^{\rm T}\left(x^{(k+N_{\rm c})}, w^{(k+N_{\rm c})}, u^{(k+N_{\rm c})}, \lambda^{(k+N_{\rm c})},  \mu^{(k+N_{\rm c})}\right)}{\partial u} \\ \\
\frac{\partial H^{\rm T}\left(x^{(k+N_{\rm c})}, w^{(k+N_{\rm c})}, u^{(k+N_{\rm c})}, \lambda^{(k+N_{\rm c})},  \mu^{(k+N_{\rm c})}\right)}{\partial \lambda}\\ \\
\frac{\partial H^{\rm T}\left(x^{(k+N_{\rm c})}, w^{(k+N_{\rm c})}, u^{(k+N_{\rm c})}, \lambda^{(k+N_{\rm c})},  \mu^{(k+N_{\rm c})}\right)}{\partial \mu}\\
\vdots\\
\frac{\partial H^{\rm T}\left(x^{(k+N_{\rm p})}, w^{(k+N_{\rm p})}, u^{(k+N_{\rm c})}, \lambda^{(k+N_{\rm p})},  \mu^{(k+N_{\rm p})}\right)}{\partial u} \\ \\
\frac{\partial H^{\rm T}\left(x^{(k+N_{\rm p})}, w^{(k+N_{\rm p})}, u^{(k+N_{\rm c})}, \lambda^{(k+N_{\rm p})},  \mu^{(k+N_{\rm p})}\right)}{\partial \lambda}\\ \\
\frac{\partial H^{\rm T}\left(x^{(k+N_{\rm p})}, w^{(k+N_{\rm p})}, u^{(k+N_{\rm c})}, \lambda^{(k+N_{\rm p})},  \mu^{(k+N_{\rm p})}\right)}{\partial \mu}
\end{matrix} \right ]=0\label{F_eq}
\end{align}

\subsection{Continuation method}
 To solve $F\left(X,U,t\right) = 0$ with respect to the unknown vector $U$, for each time-step, the C/GMRES method is employed \cite{ohtsuka2004continuation}.
In C/GMRES method, instead of solving $F(X,U,t) = 0$, we select the proper initial value $U(0)$ and take the time derivative of Eq. (\ref{F_eq}) into account.  Specifically, we define 
\begin{align}
    \dot{F}\left(X,U,t\right)=A_{\rm s}F\left(X,U,t\right)
    \label{F_stabl}
\end{align}
where $A_{\rm s}$ is a stable matrix (i.e. with negative eigenvalues). Differentiating  the left side of Eq. (\ref{F_stabl}) yields
\begin{align}
    {F}_U\left(X,U,T\right)\dot{U}=A_{\rm s}F\left(X,\ U,\ t\right)-{{F}}_X\left(X,U,T\right)\dot{X}-\dot{F}\left(X,U,T\right) 
\end{align}
If $F_U$ is non singular, we can obtain the differential equation for $\dot{U}$ as
\begin{align}
    \dot U= F^{-1}_U \left(A_{\rm s}F\left(X,\ U,\ t\right)-{{F}}_X\left(X,U,T\right)\dot{X}-\dot{F}\left(X,U,T\right)\right) \label{U_dot_QE}
\end{align}

\subsection{Forward difference GMRES method}

The calculation of Jacobians $F_x, F_U$ and $\dot F$ is computationally expensive. Instead to solve Eq. (\ref{U_dot_QE}),  we employed the forward-difference approximation to eliminate the calculation of the Jacobians. To this end,  using the concept of forward difference, we approximate the products of Jacobians and some $L \in \mathbb{R}^{11 \times N_{\rm p}}$, $M \in \mathbb{R}^{12 \times N_{\rm p}}$, and $\omega \in \mathbb{R}$ and replaced it to Eq. (\ref{U_dot_QE}) which results in:
\begin{align}
    &D_{\rm h}F\left(X,U,t:0,\dot{U},0\right)= b\left(X,\dot{X},U,t\right) \label{Fist_order_EQ}
    \end{align}
    where 
    \begin{align}
    &b\left(X,\dot{X},U,t\right) =  A_{\rm s}F\left(X,U,t\right)-D_{\rm h}F\left(X,U,t:\dot{X},0,1\right) \\
    &D_{\rm h}F\left(X,U,t:L,M,\omega\right)=\frac{F\left(X+hL,U+hM,t+h\omega \right)-F\left(X,U,t\right)}{h}
\end{align}
where $h$ is a positive real number, $D_{\rm h}F\left(X,U,t:L,M,\omega\right)$ stands for the concept of forward difference for $F$. 
It should be noted that there is main difference between forward-difference approximation and finite-difference approximation with regards to computational expenses. The forward difference approximation of the products of the Jacobians and vectors can be calculated with only an additional evaluation of the function, which requires notably less computational burden than approximation of the Jacobians themselves. Since Eq. (\ref{Fist_order_EQ}) is a linear equation with respect to $\dot{U}$, we applied the forward difference GMRES method to solve it \cite{kelley1995iterative}. The details of this method is described in Algorithm \ref{FDGMRES_Algo}.

%

 \subsection{Combination of continuation and GMRES}

  $\dot{U}$ is the output of the forward-difference GMRES algorithm, and integration of this value results in $U$ for the current time step.  For a sampling time  $\bigtriangleup t$ and integer value $\ell$,  the continuation/GMRES method for nonlinear model predictive control is summarized as follow:

\begin{algorithm}[H]
\SetAlgoLined
\KwResult{${U}$ := CntFDGMRES$\left(X,\dot{X},\dot{U},t, \bigtriangleup t, \delta\right)$}
 \do {(1) \ $t:= 0,\ \ell:=0$};\\
 (2) \ Select small value $\delta > 0$;\\
 (3)\ Find $U(0)$ for satisfying $\|F\left(X(0),U(0),0 \right)\|\leqslant \delta$ ;\\
 (4) \ In $t' \in[t, t+\bigtriangleup t)$ set $u(t'):=P_0 U(\ell\bigtriangleup t)$;\\
 At time $t+\bigtriangleup t$ by considering measured states $x(t+\bigtriangleup t)$ set $\bigtriangleup x_\ell= x_\ell(t+\bigtriangleup t) - x_\ell(t)$;\\
 (5) \  $U_{\rm int} =U_{t_0}$, $U_{\rm int} = \dot{U}((\ell-1)\bigtriangleup t)$;\\
 (6) \ $\dot{U}(\ell \bigtriangleup t)$ :=  FDGMRES$\left(X,\bigtriangleup x_\ell/\bigtriangleup t,U,U_{\rm int},h,\mathcal I_{\max}\right)$;\\
 (7) \ Set ${U}\left (\left ( \ell+1 \right ) \bigtriangleup t \right )= {U}\left (\ell \bigtriangleup t \right )+\bigtriangleup t\ \dot{U}\left (\ell \bigtriangleup t \right )$\\
 (8) \ Set $t:= t+\bigtriangleup t, \ \ell:= \ell+1$ and go back to line (4)
 \caption{Continuation/GMRES \cite{ohtsuka2004continuation}}
 \label{C_GMRES_ALG}
\end{algorithm}

It should be noted that the C/GMRES is an iterative method that solves Eq.(\ref{F_eq}) with respect to $\dot U$ only once at each sampling time and therefore, requires much less computational expenses than other iterative methods such as Newton’s method. Moreover, C/GMRES involves no line search, which is also a significant difference from standard optimization methods \cite{ohtsuka2004continuation}.

\begin{figure}[htbp]
    \centering
    \includegraphics[width=1\textwidth]{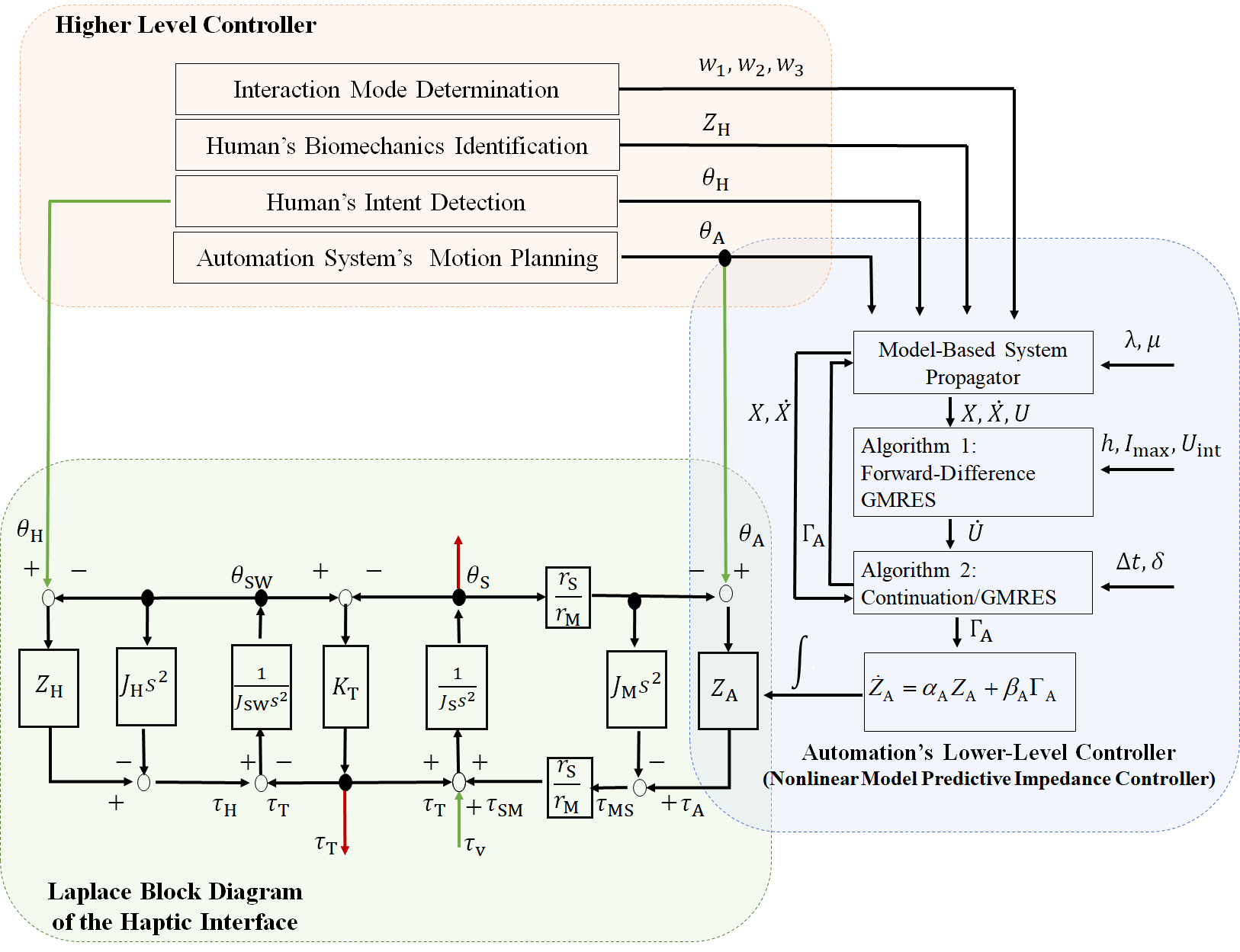}
    \caption{The detailed block diagram of the adaptive haptic shared control paradigm including the higher level controller, the automation's lower-level controller and Laplace block diagram of the haptic interface.}
    \label{Sim_Block}
\end{figure}

Figure \ref{Sim_Block} shows the control architecture of the
closed loop. The higher-level control consists of four main sections: interaction mode determination that define the appropriate form of the cost function as well as the appropriate  weights of each term in the cost function,  human's biomechanics identification that identifies the current state of $Z_{\rm H}$, human's intent detection that determines $\theta_{\rm H}$ and automation system's motion planning that determine $\theta_{\rm A}$. The outputs of the higher-level controller are fed to the automation's lower-level control to determine the optimal $Z_{\rm A}$. By modulating the automation's impedance controller gains, the control authority dynamically exchanges between the human and automation, and subsequently $\theta_{\rm S}$ follows the intents of humans or automation. 

\section{Simulation Studies and Discussions}
In this section,  we present a series of simulation studies demonstrating the effectiveness of the proposed controller in transferring the control authority between the driver and the automation system.  The following simulations consider a scenario where the human and the automation system detect an obstacle and negotiate on controlling the steering wheel so that the obstacle can be avoided safely. To this end, the simulations involve two conditions when the control authority shifts from the human to the automation system (active safety mode), as well as when the control authority shifts from the automation system to human (auto-pilot mode). Also, we included two other conditions where the human and automation are in cooperative and uncooperative mode. In cooperative mode, the human and automation intents detect the obstacle and decide to avoid the obstacle by maneuvering in the same direction (same intent signs ${\rm sgn}(\theta_{\rm H}) = {\rm sgn} (\theta_{\rm A})$).  In uncooperative mode, humans and automation's detect the obstacle but their intents have opposite signs (${\rm sgn}(\theta_{\rm H}) = -{\rm sgn} (\theta_{\rm A})$). Additionally, we assume no feedback from the road and consider $\tau_{\rm V}=0$ in the following examples. The numerical values for the parameters in the simulation are demonstrated in table (\ref{Num_param}).

\begin{table}[htbp]
    \centering
    \small
    \begin{tabular}{|c|c|c|c|c|}
\hline \multirow{2}{*}{Parameters} & \multirow{2}{*}{Variables}  & \multicolumn{2}{|c|}{Interaction Modes} & \multirow{2}{*}{Units} \\ \cline{3-4}
 &  &  Active safety & Autopilot  &  \\ 
  \hline  Activation coefficient of $k_{\rm A}$   &  $\beta_{\rm k_{\rm A}}$ &  1 & 0.1  &- \\
  \hline  Activation coefficient of $b_{\rm A}$    &  $\beta_{\rm b_{\rm A}}$ &  1 & 0.1  &- \\
   \hline  Memory coefficient of $k_{\rm A}$   &  $\alpha_{\rm k_{\rm A}}$  & \multicolumn{2}{|c|}{$-1$}& - \\
\hline  Memory coefficient of $b_{\rm A}$    &  $\alpha_{\rm b_{\rm A}}$  & \multicolumn{2}{|c|}{$-1$}& -\\
  \hline  Driver arm's stiffness   &  $K_{\rm H}$ &  0.1 & 1  & N.m/rad\\
 \hline   Driver arm's damping   &   $B_{\rm H}$ &0.1 & 0.5 &  N.m.s/rad\\
\hline   Driver arm's inertia    &  $J_{\rm H}$ & \multicolumn{2}{|c|}{$1 \times 10^{-3}$}& kg.m$^2$\\
 \hline Steering wheel inertia    & $J_{\rm SW}$ & \multicolumn{2}{|c|}{$1 \times 10^{-2}$}& kg.m$^2$\\
 \hline Steering column inertia       & $J_{\rm S}$ & \multicolumn{2}{|c|}{$1 \times 10^{-2}$}& kg.m$^2$ \\
 \hline Motor's  inertia        & $J_{\rm M}$  & \multicolumn{2}{|c|}{$1 \times 10^{-3}$}& kg.m$^2$ \\
 \hline Torque sensor stiffness         & $K_{\rm T}$ & \multicolumn{2}{|c|}{$1000$} & N.m/rad\\
 \hline Timing belt mechanical advantage       & $r_{\rm S}/r_{\rm M}$ & \multicolumn{2}{|c|}{1} & -\\
 \hline  Prediction horizon      & $N_{\rm p}$ & \multicolumn{2}{|c|}{$10$} & - \\
 \hline  Control horizon      & $N_{\rm c}$ & \multicolumn{2}{|c|}{$10$} & - \\
  \hline The sample time      & $T_{\rm S}$ & \multicolumn{2}{|c|}{$1 \times 10^{-2}$}& sec\\
  \hline Maximum index      & $\mathcal I_{\rm max}$ &  \multicolumn{2}{|c|}{$12$}& - \\
   \hline KKT vector norm range    & $\delta$&  \multicolumn{2}{|c|}{$5 \times 10^{-2}$}& -\\
  \hline
    \end{tabular}
    \caption{ Numerical values for the system parameters in the simulation}
    \label{Num_param}
\end{table}

{Figure \ref{oneagent} demonstrates the steering angle and the measured torque $\tau_{\rm T}$ for a case when there is no automation system, and the driver acts alone.  The driver's intent is expressed by the following curve
\begin{align}
    \theta_{\rm H}=\begin{cases}
    0 \hspace {105 pt} &t<T_1  \\
    \frac{W}{2}\cos({\frac{\pi}{T_2}t-\frac{T_1+T_2}{T_2}\pi}) + \frac{W}{2} \hspace {10 pt}  &T_1<t<T_1+T_2\\
W \hspace {10 pt}  &T_1+T_2<t<T_1+T_2+T_3\\
\frac{W}{2}\cos({\frac{\pi}{T_2}t-\frac{T_1+T_2+T_3}{T_2}\pi}) + \frac{W}{2} \hspace {10 pt}  &T_1+T_2+T_3<t<T_1+2T_2+T_3\\
0\hspace {10 pt}  &T_1+2T_2+T_3<t
    \end{cases}.
\end{align}
where $T_1=1$ sec, $T_2=5$ sec, $T_3=2$ sec and $W=1$ rad are selected for the following examples. Note that in the following examples to illustrate the results clearly, we select $|\theta_{\rm A}|=0.9 |\theta_{\rm H}|$.

It follows from Figure \ref{oneagent} that as the driver increases its stiffness and damping, the steering performance also increases. In particular, for three cases demonstrated in Figure \ref{oneagent},  the mean $\mu_{e}$ and the stand deviation $\sigma_e$ of the performance error $e=|\theta_{\rm H}-\theta_{\rm S}|$ are shown in table \ref{tableerror}.

\begin{table}[htbp]
    \centering
    \begin{tabular}{||c|c|c||}
    \hline
       &   $\mu_{e}$ & $\sigma_{e}$  \\\hline 
        $Z_{\rm H}=[0.1 \ 0.1]^{\rm T} $ &    0.2327
&   0.1949 \\
        $Z_{\rm H}=[0.3 \ 0.5]^{\rm T} $   & 0.0777
&    0.0651 \\
        $Z_{\rm H}=[0.5 \ 1]^{\rm T} $   & 0.0426
&    0.0358 
\\\hline
    \end{tabular}
    \caption{The mean and stand deviation of the  driver's steering performance error for different values of $Z_{\rm H}$}
    \label{tableerror}.
\end{table}

\begin{figure}[htbp]
    \centering
    \includegraphics[width=1\textwidth]{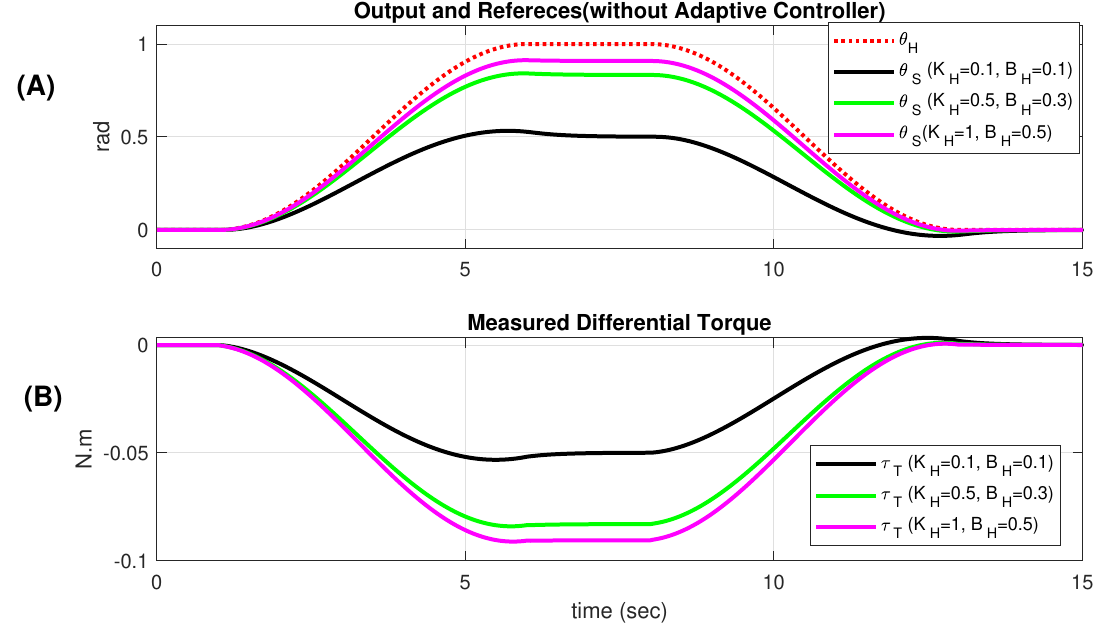}
    \caption{The driver's steering performance for different values of $Z_{\rm H}$. By increasing $Z_{\rm H}$, the driver's torque $\tau_{\rm H} \approx \tau_{\rm T}$ increases and the steering performance error $e=|\theta_{\rm H}- \theta_{\rm S}|$ decreases. }
    \label{oneagent}
\end{figure}
}

 In the following simulations, we select the human arms' bio-mechanics as either $Z_{\rm H}=[0.5\ 1]^{\rm T}$ representing a case when the human control command is sufficient or $Z_{\rm H}=[0.1 \ 0.1]^{\rm T}$ describing a situation when the driver's control command is insufficient. When the human's control command is sufficient (high $Z_{\rm H}$), the automation system is designed to yield the control authority to the human operator. Specifically, we select the weights of the cost function to be $w_1=0.2 \hat I_{3}, w_2=0_{3 \times 3}$ and $w_3=0.8 \hat I_{3}$ where $\hat I$ is the identify matrix. With selecting these weights for the cost function, the automation system acts in an auto-pilot mode \cite{bhardwaj2020s}. On the other hand, when the human's control command is insufficient (low $Z_{\rm H}$), the automation system is designed to ensure the safety of the task by avoiding the obstacle. In particular, we select the weights of the cost function to be $w_1=0_{3 \times 3}, w_2=0.8 \hat I_{3}$ and $w_3=0.2 \hat I_{3}$. With choosing these weights for the cost function, the automation system acts in the active safety mode \cite{bhardwaj2020s}.

\begin{figure}[htbp]
    \centering
    \includegraphics[width=1.1\textwidth]{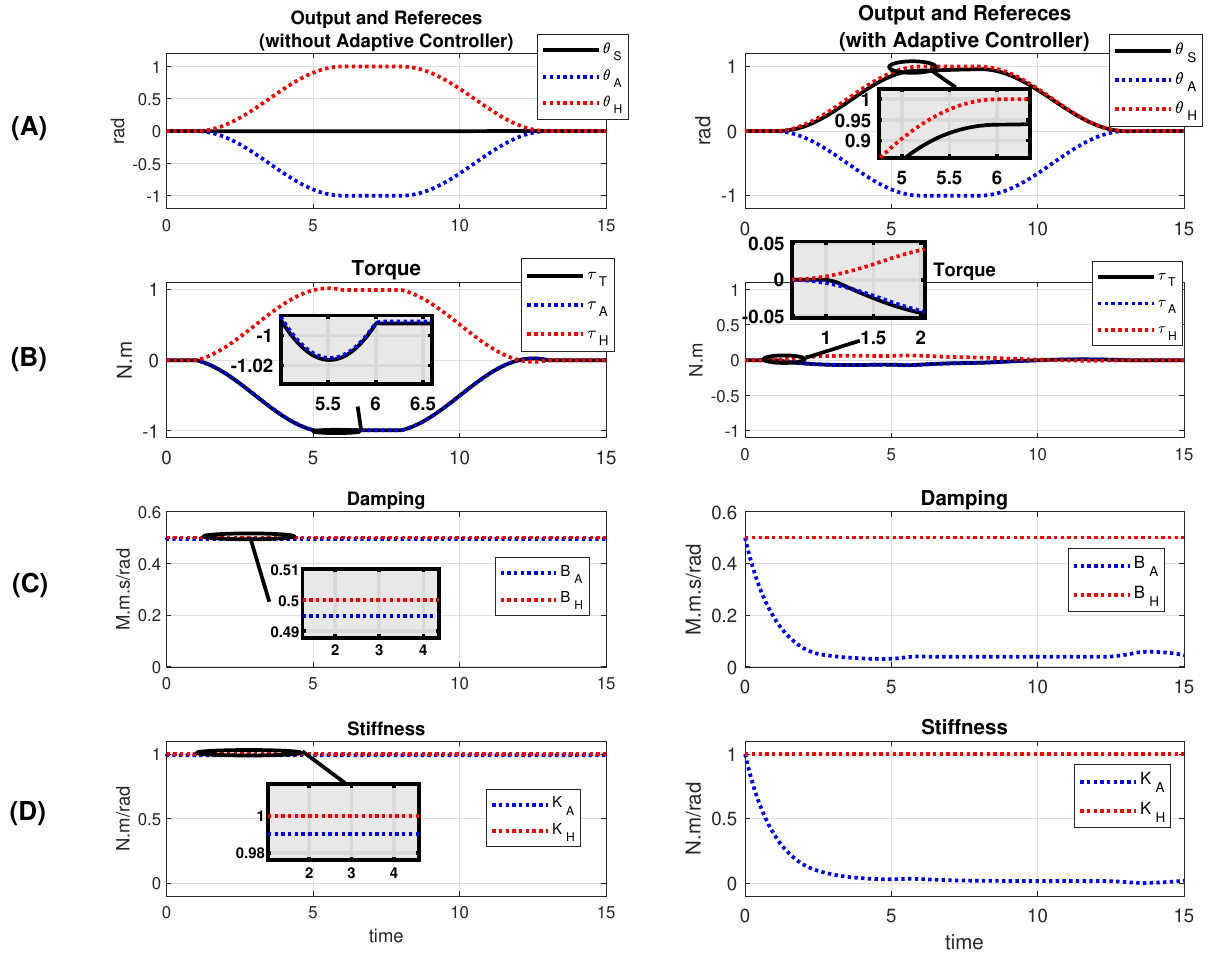}
    \caption{The outputs of the driver and automation system interaction within non-adaptive and adaptive haptic shared control paradigms are compared.  (A) driver intent (red), autonomous system intent (blue) and steering column angle (black) (B) Measured torque (black), human torque (red) and automation torque (blue) (C) Damping coefficients of the agents (D) Stiffness coefficients of the agents. The automation system act as autopilot in an uncooperative mode in the adaptive haptic shared control paradigm. By reducing the automation's impedance controller gains, the automation system reduces the disagreement between the human and automation system.}
    \label{noncop-autopilot}
\end{figure}

Figure \ref{noncop-autopilot} demonstrates the problem of control authority negotiation in un-cooperative mode. Specifically, the interaction between the human and automation system in the adaptive haptic shared control paradigm is compared with the interaction in non-adaptive haptic shared control wherein the parameters of the automation's impedance controller are invariant.  The first row shows the human's intent $\theta_{\rm H}$, the automation's intent $\theta_{\rm A}$, and the steering angle $\theta_{\rm S}$. The second row shows the human's torque $\tau_{\rm H}$, the automation system's torque $\tau_{\rm A}$, and the torque measured by the torque sensor $\tau_{\rm T}$. The third and fourth row shows the parameters of the damping and stiffness of the human arm and automation's impedance controller, respectively.  In this example, we select the human's biomechanics to be $Z_{\rm H}=[0.5 \ 1]^{\rm T}$.  It follows from Figure \ref{oneagent} that with $Z_{\rm H}=[0.5 \ 1]^{\rm T}$, the human's control command is sufficient to maneuver the steering angle safely. Therefore, we select the weights of the cost function such that the automation acts in an auto-pilot mode \cite{bhardwaj2020s} (i.e., $w_1=0.2 \hat I_{3}, w_2=0_{3 \times 3}$ and $w_3=0.8 \hat I_{3}$).  
In this example,we assumed the automation system has an estimation of the human's biomechanics and since the human adopted a high impedance, the automation yields the control authority to the human driver. 
In a non-adaptive haptic shared control paradigm, the automation's impedance controller parameters are selected to be the same as the driver ($Z_{\rm H}=Z_{\rm A}$). It follows from Figure \ref{noncop-autopilot}-A and \ref{noncop-autopilot}-B that in the non-adaptive haptic shared control paradigm when humans and automation are in the uncooperative mode, their control commands are opposite and cancel out each other ($\tau_{\rm A}\approx-\tau_{\rm H}$); and therefore, the steering angle is almost zero ($\theta_{\rm S}\approx 0$). On the other hand, it follows from the Figures \ref{noncop-autopilot}-C and \ref{noncop-autopilot}-D that in the adaptive haptic shared control paradigm, the automation' impedance controller parameters $Z_{\rm A}$ are reduced to minimize the disagreement $\tau_{\rm T}$. It follows from Figure \ref{noncop-autopilot}-B that the disagreement between humans and automation is effectively smaller than the non-adaptive haptic shared control paradigm. Furthermore, since the human's adopted impedance is sufficient, the steering angle $\theta_{\rm S}$ command follows the human's intent $\theta_{\rm H}$. 

\begin{figure}[htbp]
    \centering
    \includegraphics[width=1.1\textwidth]{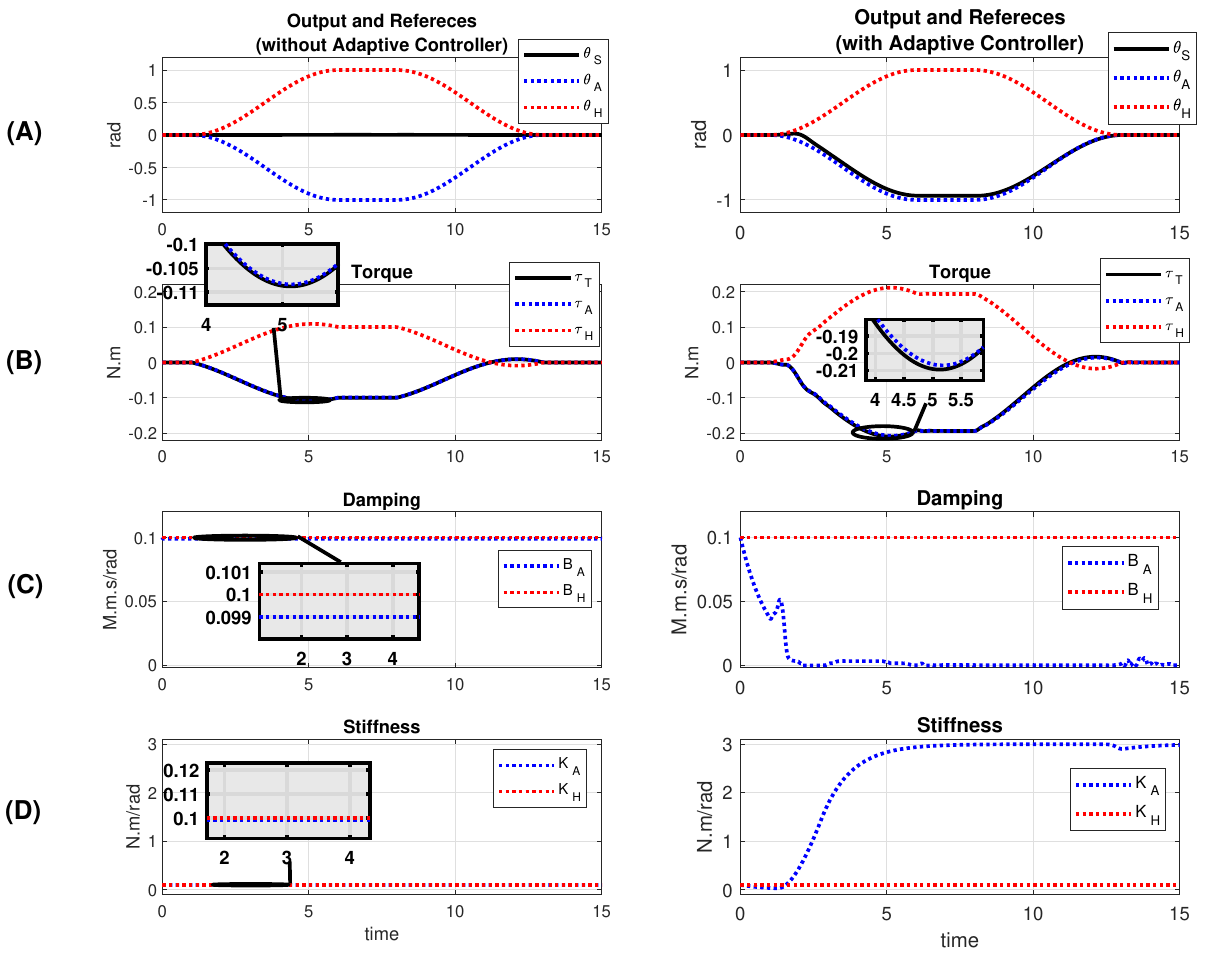}
    \caption{The outputs of the driver and automation system interaction within non-adaptive and adaptive haptic shared control paradigms are compared. (A) driver intent (red), autonomous system intent (blue) and steering column angle (black) (B) Measured torque (black), human torque (red) and automation torque (blue) (C) Damping coefficients of the agents (D) Stiffness coefficients of the agents. The automation system act as autopilot in an uncooperative mode in the adaptive haptic shared control paradigm. The automation system provides enough control input for obstacle avoidance by increasing the automation's impedance controller gains.  }
    \label{noncop-activesafety}
\end{figure}

Figure \ref{noncop-activesafety} also demonstrates the interaction between the driver and the automation system in the un-cooperative mode in non-adaptive and adaptive haptic shared control paradigms.   In this example, we selected the human's biomechanics to be $Z_{\rm H}=[0.1 \ 0.1]^{\rm T}$.  It follows from Figure \ref{oneagent} that with $Z_{\rm H}=[0.1 \ 0.1]^{\rm T}$, the human's control command is insufficient to maneuver the steering angle safely. Therefore, we select the weights of the cost function such that the automation acts in an active safety mode \cite{bhardwaj2020s} (i.e., $w_1=0_{3 \times 3}, w_2=0.8 \hat I_{3}$ and $w_3=0.2 \hat I_{3}$). Similar to the previous example, we assumed the automation system has an estimation of the human's biomechanics and since the human adopted a lower impedance, the automation re-gain the control authority from the human driver.
Also, similar to the previous example, in the non-adaptive paradigm,  the automation's impedance controller parameters are selected to be the same as the driver, and therefore,  the control commands of the human $\tau_{\rm H}$ and automation system $\tau_{\rm A}$ are opposite and cancel out each other (see Figure \ref{noncop-activesafety}-A). On the other hand, it follows from the Figures \ref{noncop-activesafety}-C and \ref{noncop-activesafety}-D that in the adaptive haptic shared control paradigm, the automation system's impedance controller parameters $Z_{\rm A}$ is increased to ensure the desired performance (e.g., avoiding an obstacle in the middle of the road). Since the automation's impedance control parameters are increased, the disagreement $\tau_{\rm T}$ between the two agents is also increased (See Figure \ref{noncop-activesafety}-B). Furthermore, since $Z_{\rm A}$ is bigger than $Z_{\rm H}$, the steering angle $\theta_{\rm S}$ is closer to the automation's intent $\theta_{\rm A}$ (see Figure \ref{noncop-activesafety}-A). 

\begin{figure}[htbp]
    \centering
    \includegraphics[width=1.1\textwidth]{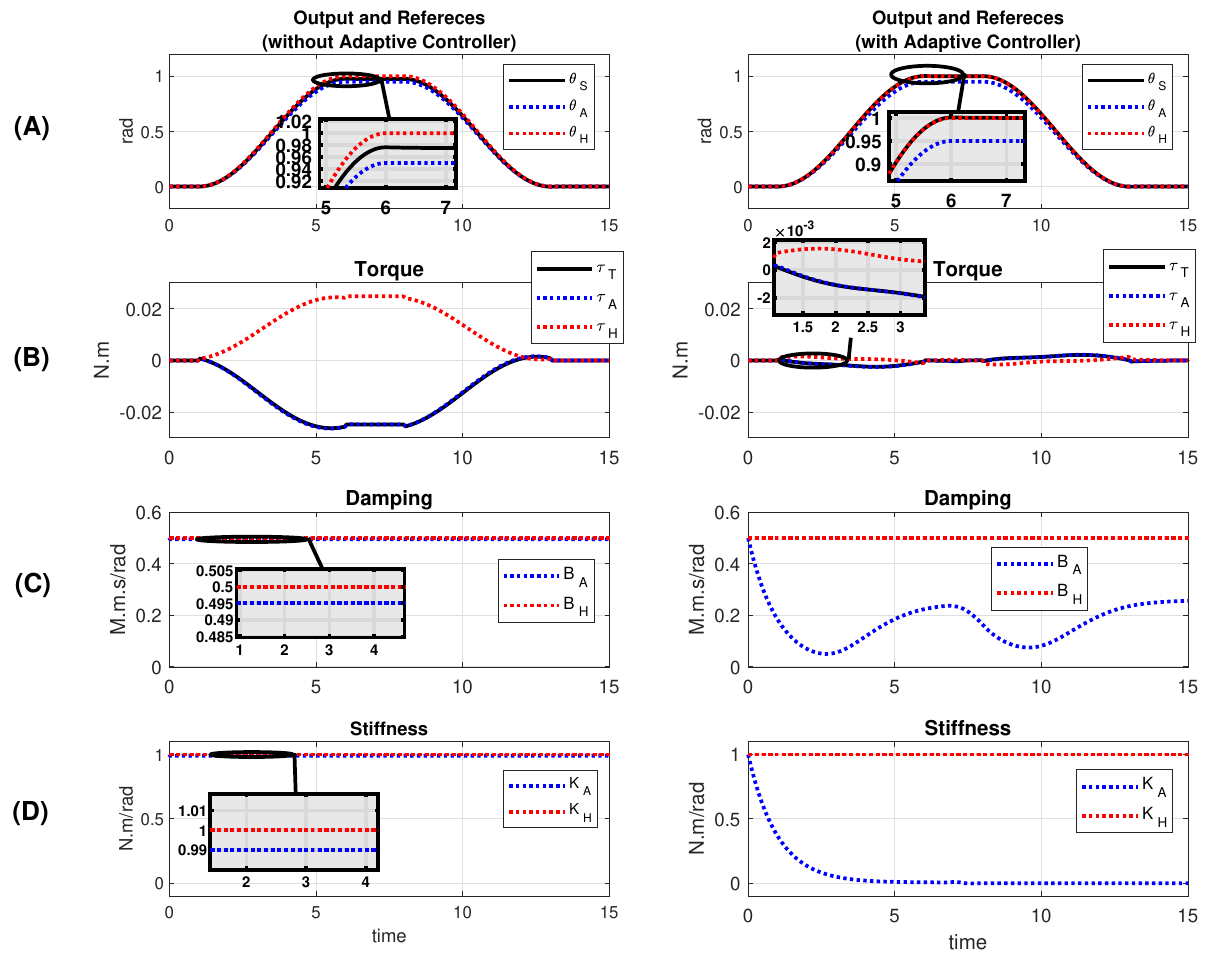}
    \caption{The outputs of the driver and automation system interaction within non-adaptive and adaptive haptic shared control paradigms are compared.  (A) driver intent (red), autonomous system intent (blue) and steering column angle (black) (B) Measured torque (black), human torque (red) and automation torque (blue) (C) Damping coefficients of the agents (D) Stiffness coefficients of the agents. 
    The automation system act as autopilot in a cooperative mode in the adaptive haptic shared control paradigm. By reducing the automation's impedance controller gains, the automation system reduces the disagreement between the human and automation system. }
    \label{coop-autopilot}
\end{figure}

Figure \ref{coop-autopilot} demonstrates the interaction between the driver and the automation system in the cooperative mode. In this example, we select the human's biomechanics to be $Z_{\rm H}=[0.5 \ 1]^{\rm T}$.  Since the human's control command is sufficient to maneuver the steering wheel safely, we select the weights of the cost function as $w_1=0.2 \hat I_{3}, w_2=0_{3 \times 3}$ and $w_3=0.8 \hat I_{3}$. With choosing these weights for the cost function, the automation system acts in an auto-pilot mode \cite{bhardwaj2020s}. In a non-adaptive haptic shared control paradigm, the automation's impedance controller parameters are selected to be the same as the driver $(Z_{\rm A}=Z_{\rm H})$. Although the torques of the driver $\tau_{\rm H}$ and the automation system $\tau_{\rm A}$ in cooperative mode are much smaller than torques in the uncooperative mode,  it follows from the Figures \ref{coop-autopilot}-B that by modulating the automation' impedance controller parameters $Z_{\rm A}$  the disagreement $\tau_{\rm T}$ even decreased more.

\begin{figure}[htbp]
    \centering
    \includegraphics[width=1.1\textwidth]{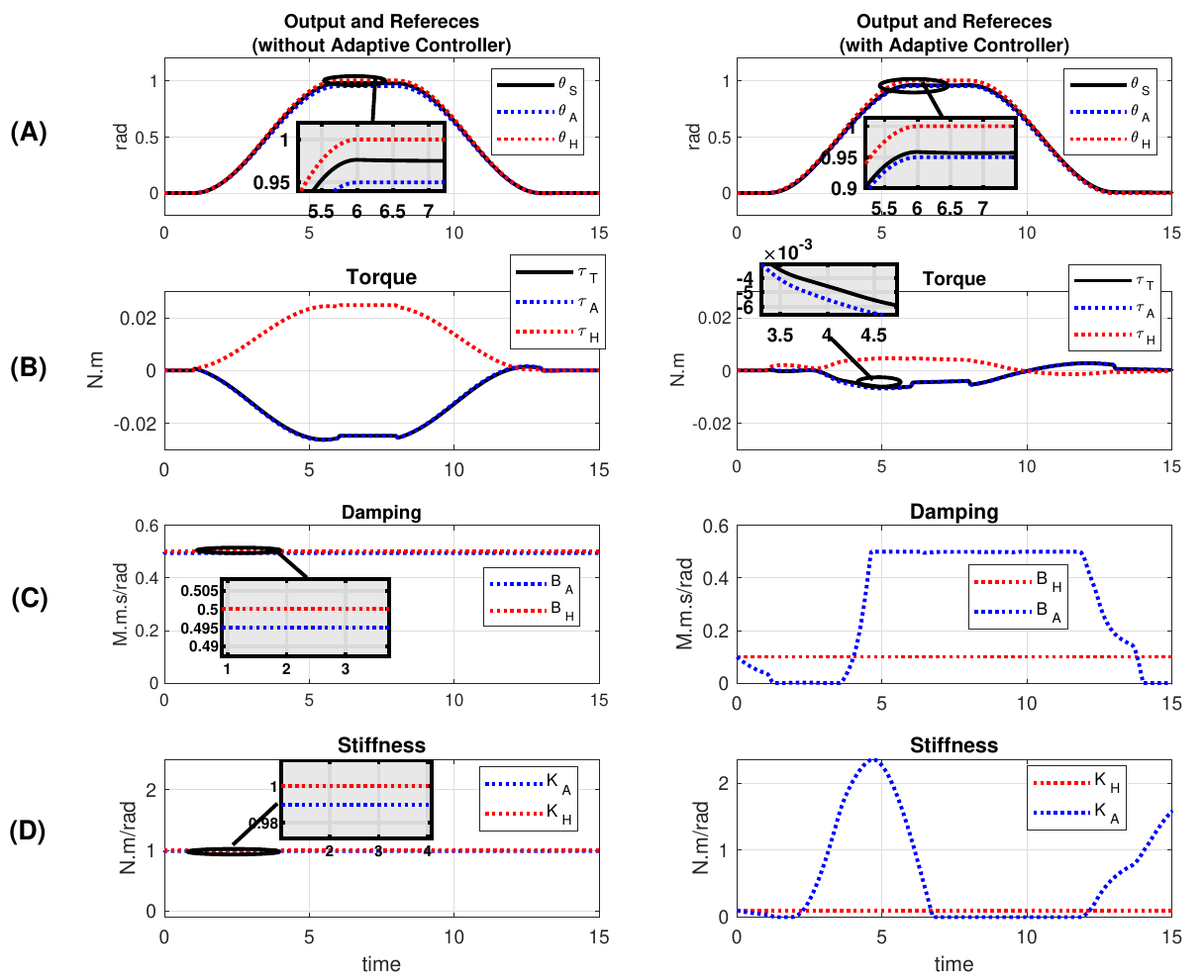}
    \caption{he outputs of the driver and automation system interaction within non-adaptive and adaptive haptic shared control paradigms are compared. (A) driver intent (red), autonomous system intent (blue) and steering column angle (black) (B) Measured torque (black), human torque (red) and automation torque (blue) (C) Damping coefficients of the agents (D) Stiffness coefficients of the agents.  The automation system act as active safety in a cooperative mode  in  the  adaptive  haptic  shared  control  paradigm. By increasing the automation's impedance controller gains, the automation system provides enough control input for the obstacle avoidance. }
    \label{coop-activesaftey}
\end{figure}

Figure \ref{coop-activesaftey} also demonstrates the interaction between the driver and the automation system in non-adaptive and adaptive haptic shared control paradigms in the cooperative mode.   In this example, we select the human's biomechanics to be $Z_{\rm H}=[0.1 \ 0.1]^{\rm T}$.  Since the human's control command is insufficient to maneuver the steering wheel safely, we select the weights of the cost function to be $w_1=0_{3 \times 3}, w_2=0.8 \hat I_{3}$ and $w_3=0.2 \hat I_{3}$. With choosing these weights for the cost function, the automation system acts in the active safety mode \cite{bhardwaj2020s}.  It follows from the Figures \ref{coop-activesaftey}-C and \ref{coop-activesaftey}-D that in the adaptive haptic shared control paradigm, the automation' impedance controller parameters $Z_{\rm A}$ are increased to ensure the desired performance (e.g., providing the required control inputs). 

\begin{figure}[htbp]
    \centering
    \includegraphics[width=1\textwidth]{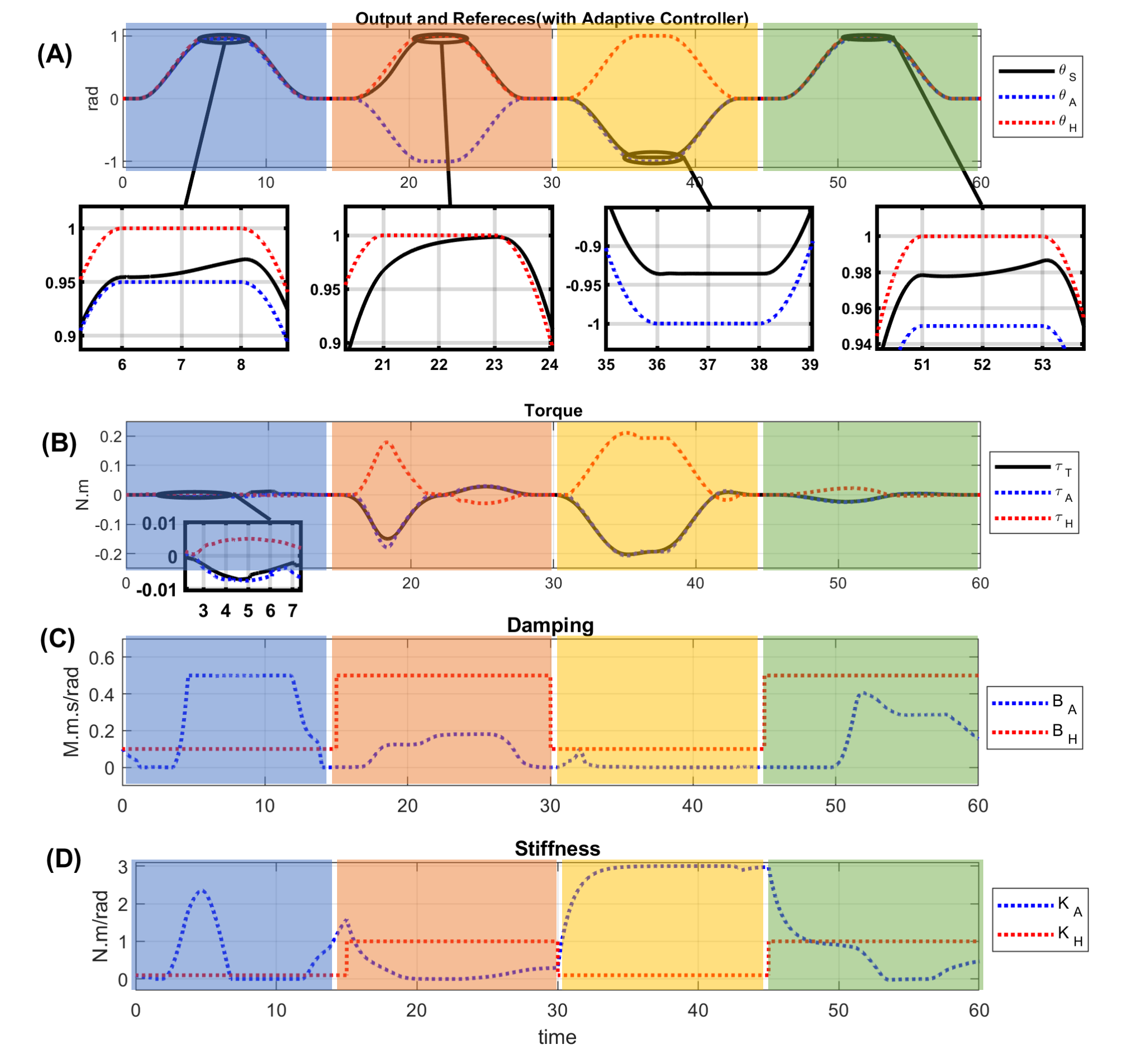}
    \caption{The human and automation's interaction in the four interaction modes. The sequence of these interaction modes is cooperative-active safety (shaded blue), uncooperative autopilot (shaded orange), uncooperative-active safety (shaded yellow), and cooperative-auto pilot mode (shaded green). The outputs of the driver and automation system interaction within non-adaptive and adaptive haptic shared control paradigms are compared. (A) driver intent (red), autonomous system intent (blue) and steering column angle (black) (B) Measured torque (black), human torque (red) and automation torque (blue) (C) Damping coefficients of the agents (D) Stiffness coefficients of the agents. In the proposed adaptive haptic shared paradigm,  by recognizing the interaction mode,  the appropriate set of weights for the cost function can be determined, and automation can continuously adjust its impedance controller parameters such that not only the safety is ensured, but also the customizability feature of the automation system is improved.
}
    \label{combined}
\end{figure}

Figure \ref{combined} shows a scenario wherein all the four interaction modes are integrated into one unified framework. The sequence of these interaction modes is cooperative-active safety, uncooperative autopilot, uncooperative-active safety, and cooperative-auto pilot mode. It follows from Figure \ref{combined} that initially, the human and automation system are in cooperative mode;  however, the human's torque input is insufficient (low $Z_{\rm H}$). The automation system increases its impedance to provide the required control command. In the next mode, the human and robot are in the uncooperative mode; however, the human's torque input is sufficient (high $Z_{\rm H}$). The automation system reduces its impedance to minimize the disagreement with the driver. In the third mode, the human and robot are in the uncooperative mode; however, the human's torque input is insufficient (low $Z_{\rm H}$). The automation system again increases its impedance to ensure safety at the expense of fighting with the driver (high $\tau_{\rm T}$).  Finally, the human and robot are again in the cooperative mode in the fourth mode; however, the human's torque input is sufficient (high $Z_{\rm H}$). The automation system reduces its impedance and yields the control authority to the driver. It follows from Figure \ref{combined} that in the proposed adaptive haptic shared paradigm,  by recognizing the interaction mode,  the appropriate set of weights for the cost function can be determined and automation can continuously adjust its impedance controller parameters such that not only the safety is ensured, but also the customizability feature of the automation system is improved.

\section{Conclusions and Future Studies}
In this paper, an adaptive haptic shared control paradigm is modeled wherein the human and automation system each is modeled as an agent with a two-level hierarchical control approach. To allow the co-robot to dynamically and continuously negotiate the control authority with the driver, we developed an optimal control approach. Specifically, we employed a nonlinear model predictive control to determine the optimal values of the automation's impedance controller. To solve the nonlinear control approach, we used the Continuation-GMRES method. A series of numerical simulations are conducted to demonstrate the effectiveness of the adaptive haptic shared control paradigm in negotiating the control authority. Our simulations involved two conditions when the control authority shifts from the human to the automation system (active safety mode), as well as when the control authority shifts from the automation system to human (auto-pilot mode). Also, we included two scenarios where the human and automation are in cooperative and uncooperative mode.

A set of challenges needs to be addressed prior to implementing the proposed shared control paradigm in real-world applications. These challenges are the subjects of our future studies. It follows from Figure \ref{Sim_Block};  the current state of the human partner's bio-mechanics is assumed to be known. Therefore, it is essential to develop an approach to estimate the parameters of human biomechanics and track them in real-time as they vary. Furthermore, it is crucial to create a method that allows recognizing the current interaction mode in real-time using the data acquired by on-board sensors \cite{kucukyilmaz2019online}. By knowing the interaction mode, an appropriate cost function can be defined, and the automation system can adjust its behavior based on this cost function.  Additionally, the proposed nonlinear model predictive control method should be expanded to consider uncertainty in the model parameters and the driver's behavior. Also, while the main focus of this paper has been to develop a control method for modulating the impedance controller parameters, knowing how and when to attempt transitions, is another challenge.  In this paper, we arbitrarily set the parameters of $\alpha_{\rm A}$ and $\beta_{\rm A}$. However, it is essential to test various transition schemes, including discrete and slow or fast continuous transitions, to determine an optimal speed for exchanging the control authority\cite{kucukyilmaz2013haptic}. Moreover, while this paper is focused on modulating impedance control parameters to allow a co-robot to communicate with a human partner, it is also essential to develop a model that can capture the interaction between the human and the automation system at a higher-level (intent level). Finally, the developed models shall be tested and refined in a hardware-in-the-loop test-bed.

\bibliography{Main,NRI,NRI2, NRI3}

\begin{thebibliography}{10}
\expandafter\ifx\csname url\endcsname\relax
  \def\url#1{\texttt{#1}}\fi
\expandafter\ifx\csname urlprefix\endcsname\relax\def\urlprefix{URL }\fi
\expandafter\ifx\csname href\endcsname\relax
  \def\href#1#2{#2} \def\path#1{#1}\fi

\bibitem{agah2000human}
A.~Agah, Human interactions with intelligent systems: research taxonomy,
  Computers \& Electrical Engineering 27~(1) (2000) 71--107.

\bibitem{albu2005physical}
A.~Albu-Schaffer, A.~Bicchi, G.~Boccadamo, R.~Chatila, A.~De~Luca,
  A.~De~Santis, G.~Giralt, G.~Hirzinger, V.~Lippiello, R.~Mattone, et~al.,
  Physical human-robot interaction in anthropic domains: safety and
  dependability, in: Proceeding 4th IARP/IEEE-EURON Workshop on Technical
  Challenges for Dependable Robots in Human Environments, 2005.

\bibitem{beyl2011safe}
P.~Beyl, K.~Knaepen, S.~Duerinck, M.~Van~Damme, B.~Vanderborght, R.~Meeusen,
  D.~Lefeber, Safe and compliant guidance by a powered knee exoskeleton for
  robot-assisted rehabilitation of gait, Advanced Robotics 25~(5) (2011)
  513--535.

\bibitem{boehm2016architectures}
P.~Boehm, A.~H. Ghasemi, S.~O'Modhrain, P.~Jayakumar, R.~B. Gillespie,
  Architectures for shared control of vehicle steering, IFAC-PapersOnLine
  49~(19) (2016) 639--644.

\bibitem{ghasemi2016role}
A.~H. Ghasemi, M.~Johns, B.~Garber, P.~Boehm, P.~Jayakumar, W.~Ju, R.~B.
  Gillespie, Role negotiation in a haptic shared control framework, in: Adjunct
  Proceedings of the 8th International Conference on Automotive User Interfaces
  and Interactive Vehicular Applications, ACM, 2016, pp. 179--184.

\bibitem{ghasemi2018adaptive}
A.~H. Ghasemi, H.~Rastgoftar, Adaptive haptic shared control framework using
  markov decision process, in: Dynamic Systems and Control Conference (DSCC),
  2018, ASME, 2018.

\bibitem{ghasemi2018game}
A.~H. Ghasemi, Game theoretic modeling of a steering operation in a haptic
  shared control framework, in: Dynamic Systems and Control Conference (DSCC),
  2018, ASME, 2018.

\bibitem{ghasemi2019shared}
A.~H. Ghasemi, P.~Jayakumar, R.~B. Gillespie, Shared control architectures for
  vehicle steering, Cognition, Technology \& Work (2019) 1--11.

\bibitem{vitiello2013neuroexos}
N.~Vitiello, T.~Lenzi, S.~Roccella, S.~M.~M. De~Rossi, E.~Cattin,
  F.~Giovacchini, F.~Vecchi, M.~C. Carrozza, Neuroexos: A powered elbow
  exoskeleton for physical rehabilitation, IEEE Transactions on Robotics 29~(1)
  (2013) 220--235.

\bibitem{bhardwaj2020s}
A.~Bhardwaj, A.~H. Ghasemi, Y.~Zheng, H.~Febbo, P.~Jayakumar, T.~Ersal, J.~L.
  Stein, R.~B. Gillespie, Who’s the boss? arbitrating control authority
  between a human driver and automation system, Transportation Research Part F:
  Traffic Psychology and Behaviour 68 (2020) 144--160.

\bibitem{izadi2019determination}
V.~Izadi, A.~Yeravdekar, A.~Ghasemi, Determination of roles and interaction
  modes in a haptic shared control framework, in: ASME 2019 Dynamic Systems and
  Control Conference, American Society of Mechanical Engineers Digital
  Collection, 2019.

\bibitem{gillespie1998virtual}
R.~B. Gillespie, M.~O’Modhrain, P.~Tang, D.~Zaretzky, C.~Pham, The virtual
  teacher, in: Proceedings of the ASME Dynamic Systems and Control Division,
  Vol.~64, American Society of Mechanical Engineers, 1998, pp. 171--178.

\bibitem{haanpaa1997advanced}
D.~P. Haanpaa, G.~P. Boston, An advanced haptic system for improving
  man-machine interfaces, Computers \& Graphics 21~(4) (1997) 443--449.

\bibitem{park2001virtual}
S.~Park, R.~D. Howe, D.~F. Torchiana, Virtual fixtures for robotic cardiac
  surgery, in: International Conference on Medical Image Computing and
  Computer-Assisted Intervention, Springer, 2001, pp. 1419--1420.

\bibitem{ikeura2002optimal}
R.~Ikeura, T.~Moriguchi, K.~Mizutani, Optimal variable impedance control for a
  robot and its application to lifting an object with a human, in: Proceedings.
  11th IEEE International Workshop on Robot and Human Interactive
  Communication, IEEE, 2002, pp. 500--505.

\bibitem{ikeura1997variable}
R.~Ikeura, A.~Morita, K.~Mizutani, Variable damping characteristics in carrying
  an object by two humans, in: Proceedings 6th IEEE International Workshop on
  Robot and Human Communication. RO-MAN'97 SENDAI, IEEE, 1997, pp. 130--134.

\bibitem{arai2000human}
H.~Arai, T.~Takubo, Y.~Hayashibara, K.~Tanie, Human-robot cooperative
  manipulation using a virtual nonholonomic constraint, in: Proceedings 2000
  ICRA. Millennium Conference. IEEE International Conference on Robotics and
  Automation. Symposia Proceedings (Cat. No. 00CH37065), Vol.~4, IEEE, 2000,
  pp. 4063--4069.

\bibitem{reed2008physical}
K.~B. Reed, M.~A. Peshkin, Physical collaboration of human-human and
  human-robot teams, IEEE Transactions on Haptics 1~(2) (2008) 108--120.

\bibitem{groten2009experimental}
R.~Groten, D.~Feth, H.~Goshy, A.~Peer, D.~A. Kenny, M.~Buss, Experimental
  analysis of dominance in haptic collaboration, in: Robot and Human
  Interactive Communication, 2009. RO-MAN 2009. The 18th IEEE International
  Symposium on, IEEE, 2009, pp. 723--729.

\bibitem{stefanov2010online}
N.~Stefanov, A.~Peer, M.~Buss, Online intention recognition for
  computer-assisted teleoperation, in: Robotics and Automation (ICRA), 2010
  IEEE International Conference on, IEEE, 2010, pp. 5334--5339.

\bibitem{oguz2010haptic}
S.~O. Oguz, A.~Kucukyilmaz, T.~M. Sezgin, C.~Basdogan, Haptic negotiation and
  role exchange for collaboration in virtual environments, in: Haptics
  Symposium, 2010 IEEE, IEEE, 2010, pp. 371--378.

\bibitem{oguz2012supporting}
S.~O. Oguz, A.~Kucukyilmaz, T.~M. Sezgin, C.~Basdogan, Supporting negotiation
  behavior with haptics-enabled human-computer interfaces, IEEE transactions on
  haptics 5~(3) (2012) 274--284.

\bibitem{sheridan2011adaptive}
T.~B. Sheridan, Adaptive automation, level of automation, allocation authority,
  supervisory control, and adaptive control: Distinctions and modes of
  adaptation, IEEE Transactions on Systems, Man, and Cybernetics-Part A:
  Systems and Humans 41~(4) (2011) 662--667.

\bibitem{inagaki2003adaptive}
T.~Inagaki, et~al., Adaptive automation: Sharing and trading of control,
  Handbook of cognitive task design 8 (2003) 147--169.

\bibitem{lee2014upper}
J.~Lee, A.~Ajoudani, E.~M. Hoffman, A.~Rocchi, A.~Settimi, M.~Ferrati,
  A.~Bicchi, N.~G. Tsagarakis, D.~G. Caldwell, Upper-body impedance control
  with variable stiffness for a door opening task, in: 2014 IEEE-RAS
  International Conference on Humanoid Robots, IEEE, 2014, pp. 713--719.

\bibitem{righetti2014autonomous}
L.~Righetti, M.~Kalakrishnan, P.~Pastor, J.~Binney, J.~Kelly, R.~C. Voorhies,
  G.~S. Sukhatme, S.~Schaal, An autonomous manipulation system based on force
  control and optimization, Autonomous Robots 36~(1-2) (2014) 11--30.

\bibitem{adhikary2017hybrid}
N.~Adhikary, C.~Mahanta, Hybrid impedance control of robotic manipulator using
  adaptive backstepping sliding mode controller with pid sliding surface, in:
  2017 Indian Control Conference (ICC), IEEE, 2017, pp. 391--396.

\bibitem{anderson1988hybrid}
R.~J. Anderson, M.~W. Spong, Hybrid impedance control of robotic manipulators,
  IEEE Journal on Robotics and Automation 4~(5) (1988) 549--556.

\bibitem{balatti2018self}
P.~Balatti, D.~Kanoulas, G.~F. Rigano, L.~Muratore, N.~G. Tsagarakis,
  A.~Ajoudani, A self-tuning impedance controller for autonomous robotic
  manipulation, in: 2018 IEEE/RSJ International Conference on Intelligent
  Robots and Systems (IROS), IEEE, 2018, pp. 5885--5891.

\bibitem{mori2007continuation}
H.~Mori, K.~Seki, Continuation newton-gmres power flow with linear and
  nonlinear predictors, in: 2007 Large Engineering Systems Conference on Power
  Engineering, IEEE, 2007, pp. 171--175.

\bibitem{soneda2005nonlinear}
Y.~Soneda, T.~Ohtsuka, Nonlinear moving horizon state estimation with
  continuation/generalized minimum residual method, Journal of guidance,
  control, and dynamics 28~(5) (2005) 878--884.

\bibitem{ohtsuka1994stabilized}
T.~Ohtsuka, H.~Fujii, Stabilized continuation method for solving optimal
  control problems, Journal of Guidance, Control, and Dynamics 17~(5) (1994)
  950--957.

\bibitem{ohtsuka2000continuation}
T.~Ohtsuka, Continuation/gmres method for fast algorithm of nonlinear receding
  horizon control, in: Proceedings of the 39th IEEE Conference on Decision and
  Control (Cat. No. 00CH37187), Vol.~1, IEEE, 2000, pp. 766--771.

\bibitem{kelley1995iterative}
C.~T. Kelley, Iterative methods for linear and nonlinear equations, Vol.~16,
  Siam, 1995.

\bibitem{ohtsuka2004continuation}
T.~Ohtsuka, A continuation/gmres method for fast computation of nonlinear
  receding horizon control, Automatica 40~(4) (2004) 563--574.

\bibitem{bhardwaj2019estimating}
A.~Bhardwaj, B.~Gillespie, J.~Freudenberg, Estimating rack force due to road
  slopes for electric power steering systems, in: 2019 American Control
  Conference (ACC), IEEE, 2019, pp. 328--334.

\bibitem{tapia1979role}
R.~A. Tapia, Role of slack variables in quasi-newton methods for constrained
  optimization, Tech. rep., Rice Univ., Houston, TX (USA). Dept. of
  Mathematical Sciences (1979).

\bibitem{kucukyilmaz2019online}
A.~Kucukyilmaz, I.~Issak, Online identification of interaction behaviors from
  haptic data during collaborative object transfer, IEEE Robotics and
  Automation Letters 5~(1) (2019) 96--102.

\bibitem{kucukyilmaz2013haptic}
A.~Kucukyilmaz, et~al., Haptic role allocation and intention negotiation in
  human-robot collaboration, Ph.D. thesis, Ko{\c{c}} University (2013).

\end{thebibliography}

\end{document}